\documentclass{article}
\usepackage{spconf,graphicx}
\usepackage{amsmath,amssymb}
\usepackage{url,lipsum,cite}
\usepackage{multicol,multirow}
\usepackage{xpatch,xcolor}
\usepackage{tabularx}
\usepackage{makecell, booktabs}
\newcolumntype{Y}{>{\centering\arraybackslash}X}
\usepackage{hyperref}

\hypersetup{
    colorlinks=true,
    linkcolor=purple,
    filecolor=magenta,      
    urlcolor=purple,
}

\newcommand{\newpara}[1]{\vspace{8pt}\noindent\textbf{#1}}
\newcommand\blfootnote[1]{%
  \begingroup
  \renewcommand\thefootnote{}\footnote{#1}%
  \addtocounter{footnote}{-1}%
  \endgroup
}
\title{AASIST: Audio Anti-Spoofing using Integrated Spectro-Temporal \\Graph Attention Networks}
\name{
  \begin{tabular}{c}
  Jee-weon Jung$^1$, Hee-Soo Heo$^1$, Hemlata Tak$^2$, Hye-jin Shim$^3$, \\
   Joon Son Chung$^1$, Bong-Jin Lee$^1$, Ha-Jin Yu$^3$, Nicholas Evans$^2$
  \end{tabular}
}
\address{
  $^1$Naver Corporation, South Korea\\
  $^2$EURECOM, Sophia Antipolis, France\\
  $^3$School of Computer Science, University of Seoul, South Korea
 }
\begin{document}
\ninept
\maketitle

\begin{abstract}
Artefacts that differentiate spoofed from bona-fide utterances can reside in spectral or temporal domains.
Their reliable detection usually depends upon computationally demanding ensemble systems where each subsystem is tuned to some specific artefacts. 
We seek to develop an efficient, single system that can detect a broad range of different spoofing attacks without score-level ensembles. 
We propose a novel heterogeneous stacking graph attention layer which models artefacts spanning heterogeneous temporal and spectral domains with a heterogeneous attention mechanism and a stack node.
With a new max graph operation that involves a competitive mechanism and an extended readout scheme, our approach, named {\em AASIST}, outperforms the current state-of-the-art by 20\% relative. 
Even a lightweight variant, AASIST-L, with only 85K parameters, outperforms all competing systems.

\end{abstract}

\section{Introduction}
\label{sec:intro}
The audio anti-spoofing (i.e., spoofing detection) task can bridge a speaker verification into a real world application by determining whether an input speech utterance is genuine ({\em bona-fide}) or spoofed, thus, improving the credibility. 
Practical spoofing detection systems are required to detect spoofed utterances generated using a wide range of different techniques. 
The ASVspoof community has led research in the field with a series of challenges accompanied by public datasets~\cite{wu2015asvspoof,kinnunen2017asvspoof,todisco2019asvspoof,yamagishi2021asvspoof}.
Two major scenarios are being studied, namely logical access (LA) and physical access.
The focus in this paper is LA, which considers spoofing attacks mounted with voice conversion and text-to-speech algorithms.\blfootnote{Code and models are available at:\\\url{https://github.com/clovaai/aasist}}

Recent studies show that discriminative information (i.e., spoofing artefact) can reside in both spectral and temporal domains~\cite{yang2019significance,sriskandaraja2016investigation,jung2019replay,odyssey2020CQCC,tak2020spoofing}. 
Artefacts tend to be dependent upon the nature of the attack and the specific algorithm used.
Adaptive mechanisms, which have the flexibility to concentrate on the domain in which the artefacts lie, are therefore crucial to reliable detection. 
In our recent works~\cite{tak2021graph,tak2021end}, we proposed end-to-end systems leveraging a RawNet2-alike encoder~\cite{jung20c_interspeech,tak2021rawnet} and graph attention networks~\cite{velickovic2018graph}. 
We modeled both spectral and temporal information concurrently using two parallel graphs and then applied element-wise multiplication on two graphs. 
While we achieved state-of-the-art performance in~\cite{tak2021end}, we believe that there is still room for further improvement because two graphs are heterogeneous, integrating them using a heterogeneity-aware technique would be beneficial.  

We propose four extensions to our previous work, RawGAT-ST~\cite{tak2021end} where the first three compose the proposed model named {\em AASIST} and the last builds a smaller version of the AASIST. 
First, we propose an extended variant of the graph attention layer, referred to as a ``{\em heterogeneous stacking graph attention layer}'' (HS-GAL). 
It facilitates the concurrent modeling of heterogeneous (spectral and temporal) graph representations. 
HS-GAL includes a modified attention mechanism considering heterogeneity and an additional stack node, each inspired from~\cite{wang2019heterogeneous} and~\cite{kenton2019bert} respectively.
HS-GAL can directly model two arbitrary graphs where the two graphs can have different numbers of nodes and different dimensionalities.  
Second, we propose a mechanism referred to as ``{\em max graph operation}'' (MGO) that mimics the max feature map~\cite{wu2018light}. 
MGO involves two branches where each branch includes two HS-GALs and graph pooling layers, followed by an element-wise maximum operation. 
The underlying objective here is to enable different branches to learn different groups of artefacts. 
The elements that include artefacts would survive the maximum operation. 
Third, we present a new readout scheme that utilizes the stack node.
Finally, given the application of anti-spoofing solutions in {\em aasist}ing speaker verification systems~\cite{Kinnunen2018tdcf,shim2020integrated} and given the associated requirement for practical, lightweight models, we further propose a lightweight variant of AASIST which comprises only $85$K parameters.

\begin{figure*}[ht!]
    \centering
    \includegraphics[width=\textwidth,height=4cm]{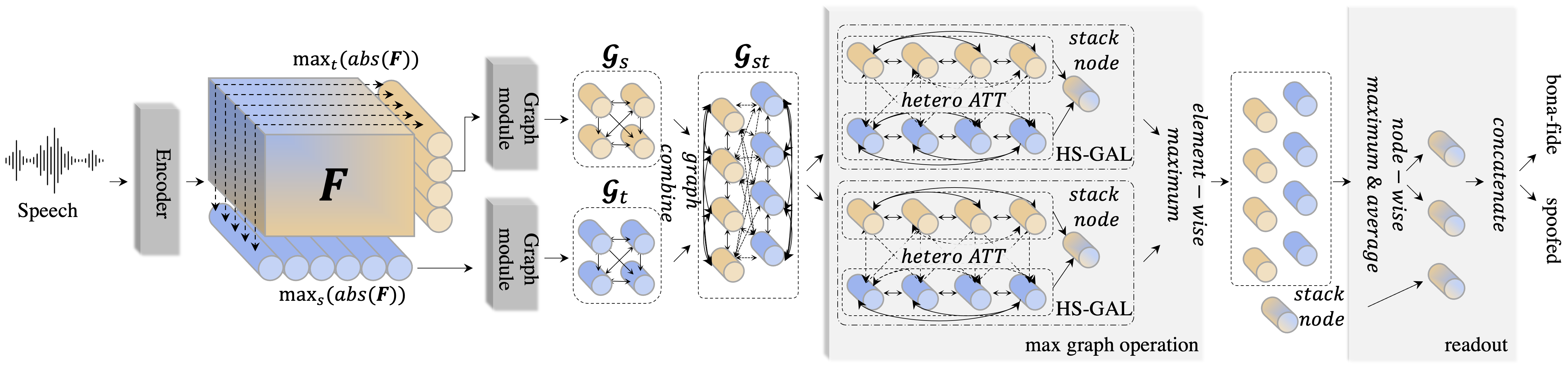}
    \caption{
      Overall framework of the proposed AASIST.
      {\em Identical to} \cite{tak2021end}: encoder extracts $F$ and two graph modules each model spectral and temporal domains. 
      {\em Proposed}: then, the proposed max graph operation technique adopts two branches that model heterogeneous graphs in parallel, followed by an element-wise maximum. 
      Each branch includes two proposed HS-GAL layers and two graph pooling layers (graph pooling layers and one HS-GAL layer is omitted in the illustration). 
      Finally, the maximum and average of nodes, and the stack node are concatenated followed by an output layer.
      }
    \vspace{-10pt}
    \label{fig:framework}
\end{figure*}

\section{Preliminaries}
In this section, we summarise the components that are common to the RawGAT-ST~\cite{tak2021end} and new AASIST models.
We describe: i)~the RawNet2-based encoder used for extracting high-level feature maps from raw input waveforms; ii)~the graph module which includes graph attention and graph pooling layers.
The two components correspond to ``encoder'' and ``graph module'' in Figure~\ref{fig:framework}, respectively.

\subsection{RawNet2-based encoder}
\label{sec:encoder}
A growing number of researchers are adopting models that operate directly upon raw waveform inputs. 
The work described in this paper utilizes a variant of the RawNet2 model introduced in~\cite{jung20c_interspeech} for the task of speaker verification and applied subsequently for anti-spoofing~\cite{tak2021rawnet,tak2021end}.
It extracts high-level representations $F$, $F\in\mathbb{R}^{C \times S \times T}$ directly from raw waveform inputs where $C, S$ and $T$ are the number of channels, spectral (frequency) bins, and the temporal sequence length respectively. 

Different to the original RawNet2 model, we interpret the output of the sinc-convolution layer as a 2-dimensional image with a single channel (akin to a spectrogram) rather than a 1-dimensional sequence with multiple filters by treating the output of each filter as a spectral bin.
A series of six residual blocks with pre-activation~\cite{he2016identity} is used to extract the high-level representation.
Each residual block comprises a batch normalization layer~\cite{ioffe2015batch}, a 2-dimensional convolution layer, SeLU activation~\cite{klambauer2017self}, and a max pooling layer. 
Further information can be found in~\cite{tak2021end}.

\subsection{Graph module}
\label{sec:GATandGPool}

\newpara{Graph attention network.}
Recent advances in graph neural networks have brought performance breakthroughs in a number of tasks~\cite{velickovic2018graph,jung2021graph,tak2021graph} where a graph is defined by a set of nodes and a set of edges connecting different node pairs. 
Using high-dimensional vector as a node, graph neural networks can be used to model the non-Euclidean data manifold between different nodes.
In particular, we have shown that the graph attention network~\cite{velickovic2018graph} can be applied to both speaker verification~\cite{jung2021graph} and spoofing detection~\cite{tak2021graph, tak2021end}.

The graph attention layer used in our work is a variant of the original architecture~\cite{velickovic2018graph}.
In our work, graphs are fully-connected in the sense of there being edges between each and every node pair. 
This is because the relevance of each node pair to the task at hand cannot be predetermined. 
Instead, the self-attention mechanism in a graph attention layer derives data-driven attention weights, assigned to each edge, to reflect the relevance of each node pair. 
Before deriving attention weights, an element-wise multiplication is utilized to make edges symmetric.
The reader is referred to~\cite{tak2021end} (Section~3) for further details. 

\newpara{Graph pooling.}
Various graph pooling layers have been proposed to effectively scale down the graph~\cite{lee2019self,gao2019graph}.
This has the aim of reducing complexity and improving discrimination.
We apply a simple attentive graph pooling layer to the output of each graph attention layer. 
Except for the omission of projection vector normalization, our implementation is identical to that in~\cite{gao2019graph}.

Let $\mathcal{G}, \mathcal{G} \in \mathbb{R}^{N \times D}$ be the output graph of a graph attention layer where $N$ is the number of nodes and $D$ refers to the dimensionality of each node. 
Note that the order of nodes is meaningless; the relationships between them are defined via the attention weights assigned to each edge. 
Attention weights are derived via $G \cdot P$ where $\cdot$ is the dot product and $P, P \in \mathbb{R}^{D}$ is a projection vector that returns a scalar attention weight for each node. 
After the multiplication of a sigmoid non-linearity with the corresponding $k$ nodes, the nodes with the top-$k$ values are retained while the rest are discarded. 

\begin{table*}[!t]
    \centering
    \small
    \setlength\tabcolsep{4.5pt}
    \begin{tabularx}{\linewidth}{l || *{13}{c}| YY}
      \Xhline{1pt}
      \textbf{System} &\textbf{A07}&\textbf{A08}&\textbf{A09}&\textbf{A10}&\textbf{A11}&\textbf{A12}&\textbf{A13}&\textbf{A14}&\textbf{A15}&\textbf{A16}&\textbf{A17}&\textbf{A18}&\textbf{A19}& \textbf{P1} & \textbf{P2 (\%)} 	\\ 
      \Xhline{1pt}
      RawGAT-ST &1.19 &\textbf{0.33} &0.03 &1.54 &0.41 &1.54 &0.14 &\textbf{0.14} &1.03 &\textbf{0.67} &\textbf{1.44} &\textbf{3.22} &\textbf{0.62} & 0.0443(0.0333) &1.39(1.19)\\
      \hline
      AASIST & \textbf{0.80} & 0.44 & \textbf{0.00} & \textbf{1.06} & \textbf{0.31} & \textbf{0.91} & \textbf{0.1} & \textbf{0.14} & \textbf{0.65} & 0.72 & 1.52 & 3.40 & \textbf{0.62} & \textbf{0.0347}(\textbf{0.0275})  &\textbf{1.13}(\textbf{0.83})\\

         \Xhline{1pt}
    \end{tabularx}
    \caption{
      Breakdown EER (\%) performance of all 13 attacks that exist in the ASVspoof 2019 LA evaluation set, pooled min t-DCF (P1), and pooled EER (\%, P2).
      RawGAT-ST~\cite{tak2021end} (baseline, state-of-the-art) and the proposed AASIST are reported. 
      We reproduced RawGAT-ST using three different random seeds where the performance of the best seed shows similar performance compared to that of the original paper (in Table~\ref{tab:sotaNsmall}).
      All reported performances are the average using three repeated experiments with different random seeds, in which values inside brackets are the performance of the best performing seed. The best performance for each column is marked in boldface. }
    \vspace{-10pt}
    \label{tab:breakdown}
\end{table*}

\section{AASIST}
\label{sec:AASIST}
AASIST builds upon the foundation of our previous work, RawGAT-ST, whereby two heterogeneous graphs, one spectral and the other temporal, are combined at the model-level. 
However, instead of using trivial element-wise operations and fully-connected layers, the new approach relies upon a more elegant approach using the proposed HS-GAL. 
Also, AASIST includes the proposed MGO and readout techniques. 

Figure~\ref{fig:framework} illustrates the overall framework of the AASIST including proposed HS-GAL, MGO, and readout techniques. 
High-level representation $F$ is extracted by feeding raw waveforms into the RawNet2-based encoder (Section~\ref{sec:encoder}). 
Two graph modules first model the spectral and the temporal domain in parallel, deriving $\mathcal{G}_s$ and $\mathcal{G}_t$ (Section~\ref{sec:GATandGPool}). 
The results are combined into $\mathcal{G}_{st}$ (Section~\ref{ssec:graphCombine}) and processed using MGO (Section~\ref{ssec:MgoReadout}) that includes four HS-GAL layers (Section~\ref{ssec:hetero}) and four graph pooling layers. 
Readout operations are then performed, followed by an output layer with two nodes.

\subsection{Graph combination}
\label{ssec:graphCombine}
We first compose a heterogeneous graph using two different graphs that each model spectral and temporal domains ($\mathcal{G}_{st}$ in Figure~\ref{fig:framework}). 
Let $\mathcal{G}_s, \mathcal{G}_s \in \mathbb{R}^{N_s \times D_s}$ and $\mathcal{G}_t, \mathcal{G}_t \in \mathbb{R}^{N_t \times D_t}$ be spectral and temporal graphs respectively, each derived according to:
\begin{align}
    \mathcal{G}_s = graph\_module(max_t(abs(F))),\\
    \mathcal{G}_t = graph\_module(max_s(abs(F))),
\end{align}
where $graph\_module$ refers to the combination of graph attention and graph pooling layers and where $F \in \mathbb{R}^{C \times S \times T}$ is the encoder output feature map.
We then formulate a combined graph $\mathcal{G}_{st}$ which has $N_s+N_t$ nodes by adding edges between every node in $G_t$ and every node in $G_s$ and vice versa (dotted arrows under $\mathcal{G}_{st}$).
The new edges in the combined graph $G_{st}$ allow deriving attention weights between pairs of heterogeneous nodes that each spans temporal and spectral domains.  
Despite the combination, $G_{st}$ remains a heterogeneous graph in that nodes in each of the constituent graphs lie in different latent spaces;  $N_s$ and $D_s$ are normally different to $N_t$ and $D_t$.   

\subsection{HS-GAL}
\label{ssec:hetero}
The new contribution is based upon a {\em heterogeneous stacking graph attention layer} (HS-GAL - dotted box in Figure~\ref{fig:framework}). 
It comprises two components, namely {\em heterogeneous attention} and a {\em stack} node.
Our approach to heterogeneous attention is inspired by the approach to the modeling of heterogeneous data described in~\cite{wang2019heterogeneous}. 

The input to the HS-GAL is first projected into another latent space to give each of the two graphs with node dimensionalities $D_t$ and $D_s$ a common dimensionality $D_{st}$.
Two fully-connected layers are utilized for this purpose, each projecting one of the constituent sub-graphs to a dimensionality of $D_{st}$.

\newpara{Heterogeneous attention.}
Whereas the homogeneous graphs use a single projection vector to derive attention weights, we use three different projection vectors to calculate attention weights for the heterogeneous graph.  
They are illustrated inside $\mathcal{G}_{st}$ of Figure~\ref{fig:framework} and are used to determine attention weights for edges connecting: 
(i)~nodes in $\mathcal{G}_{s}$ to nodes in $\mathcal{G}_{s}$ (edges between orange nodes); 
(ii)~nodes in $\mathcal{G}_{s}$ to $\mathcal{G}_{t}$ and $\mathcal{G}_{t}$ to $\mathcal{G}_{s}$ (dotted edges); 
(iii)~$\mathcal{G}_{t}$ to $\mathcal{G}_{t}$ (edges between blue nodes). 
The projection vector in the case of (ii) above applies to edges in both directions; the graph attention layer applies element-wise multiplication between two nodes, making attention weights symmetrical, rather than concatenating two nodes as in~\cite{velickovic2018graph}. 

\newpara{Stack node.}
We also introduce a new, additional node referred to as the ``{\em stack}'' node.
The role of the stack node is to accumulate heterogeneous information, namely information or the relationship between spectral and temporal domains. 
The stack node is connected to the full set of nodes (stemming from $\mathcal{G}_s$ and $\mathcal{G}_t$). 
The use of uni-directional edges from all other nodes to the stack node helps to preserve information in both $\mathcal{G}_s$ and $\mathcal{G}_t$.
It does not transmit information to other nodes. 
Also, when using more than one HS-GAL layer sequentially, the stack node of the previous layer can be passed on to the next layer. 
The behaviour of the stack node is similar to that of classification tokens~\cite{kenton2019bert}, except that connections to other nodes are uni-directional.

\subsection{Max graph operation and readout}
\label{ssec:MgoReadout}
The new ``{\em max graph operation}'' (MGO), highlighted with a large grey box in Figure~\ref{fig:framework}, is inspired by a number of works in the anti-spoofing literature which showed the benefit of element-wise maximum operations~\cite{lavrentyeva2017audio,tak2021end}. 
MGO aims to mimic the procedure of detecting various artefacts evoked by spoofing in parallel and combining them.
It utilizes two parallel branches where element-wise maximum is applied to two branches' outputs. 
Specifically, each branch involves two HS-GAL sequentially where a graph pooling layer is adopted after each HS-GAL.
Thus, MGO comprises four HS-GAL and four graph pooling layers in total. 
The two HS-GAL in each branch share the stack node by passing the stack node of preceding HS-GAL to the following HS-GAL. 
Element-wise maximum is also applied to two branches' stack nodes.

The modified readout scheme is illustrated in the right-most grey box of Figure~\ref{fig:framework}.
First, we apply node-wise maximum and average. 
The last hidden layer is then formed from the concatenation of two nodes derived using average and maximum along with the stack node. 

\subsection{Lightweight variant of AASIST}
\label{ssec:smallModel}
We additionally explore a lightweight variant, namely {\em AASIST-L}. 
Leaving the architecture identical, we tune the number of parameters to compose a model with $85$K parameters using the population-based training algorithm~\cite{jaderberg2017population}. 
This results in a $332$KB model where after techniques such as half-precision training, the size would be halved.
Through experiments, we demonstrate that AASIST-L still outperforms all models except AASIST, shown in Table~\ref{tab:sotaNsmall}. 

\begin{table*}[!t]
  \centering
   \setlength\tabcolsep{5pt}
  \begin{tabularx}{\linewidth}{lcYYYY}
    \Xhline{1pt}
	\textbf{System} & \textbf{\# Param} & \textbf{Front-end} & \textbf{Architecture} & \textbf{min t-DCF} & \textbf{EER (\%)}	\\ 
    \Xhline{1pt}
	\textbf{{\em Ours}} &\textbf{ 297K }& \textbf{Raw waveform} & \textbf{AASIST} & \textbf{0.0275} & \textbf{0.83}\\
	\hline
	\textbf{{\em Ours}} & \textbf{85K} & \textbf{Raw waveform }& \textbf{AASIST-L} & \textbf{0.0309} & \textbf{0.99}\\
	\hline
	 Tak et al.~\cite{tak2021end} & 437K & Raw waveform & RawGAT-ST & 0.0335 & 1.06\\
	 \hline
	 Zhang et at.~\cite{zhang2021effect}&1,100K&FFT&SENet&0.0368& 1.14\\
	 \hline
	 Hua et al.~\cite{hua2021towards}&350K&Raw waveform&Res-TSSDNet&0.0481&1.64\\ 
	 \hline	
	 Ge et al.~\cite{ge2021raw}&24,480K&Raw waveform &Raw PC-DARTS& 0.0517&1.77\\
	 \hline
	 Li et al.~\cite{li2021channelwise}& 960K & CQT &MCG-Res2Net50 & 0.0520&1.78\\
	  \hline	
	 Chen et al.~\cite{chen2020generalization}&-&LFB&ResNet18-LMCL-FM&0.0520&1.81\\
	 \hline
	Wang et al.~\cite{wang21fa_interspeech}&276K&LFCC&LCNN-LSTM-sum&0.0524&1.92\\  
	 \hline	
	Luo et al.~\cite{luo2021capsule}&-&LFCC&Capsule network&0.0538&1.97\\  
	\hline
	Zhang et al.~\cite{zhang2021one}&-&LFCC&Resnet18-OC-softmax&0.0590&2.19\\
	\hline
	Li et al.~\cite{li2021replay}&-&CQT&SE-Res2Net50&0.0743&2.50\\
	\hline
	Ma et al.~\cite{ma2021improved}&-&LFCC&LCNN-Dual attention&0.0777&2.76\\
	\hline
	Tak et al.~\cite{tak2021graph}&-&LFB&ResNet18-GAT-T&0.0894&4.71\\
	\hline
    Tak et al.~\cite{tak2020spoofing}&-&LFCC&GMM&0.0904 &3.50\\
	\hline
	Ge et al.~\cite{ge2021}&7,510K&LFCC& PC-DARTS& 0.0914&4.96\\
 	
    \Xhline{1pt}
	\end{tabularx}
	\caption{
	    Comparison with recently proposed state-of-the-art systems, reported using pooled min t-DCF and EER (\%). 
	    Systems are displayed in an ascending order using the min t-DCF. 
	    For the proposed AASIST and AASIST-L, we report the best single result. 
	    All results are single model performance without any kind of score-level ensemble.
	}
 	\vspace{-10pt}
	\label{tab:sotaNsmall}
\end{table*}
\begin{table}[!t]
	\centering
	\small
	\begin{tabularx}{\columnwidth}{lYY}
    \Xhline{1pt}
	 Configuration & min t-DCF& EER	\\ 
    \Xhline{1pt}
	 AASIST & 0.0347(0.0275) & 1.13(0.83)\\
	 \hline\hline
	 w/o heterogeneous attention & 0.0415(0.0384) & 1.44(1.37)\\
	 \hline
	 w/o stack node & 0.0380(0.0330) & 1.21(1.03)\\
	 \hline
	 w/o MGO & 0.0410(0.0378) & 1.35(1.19)\\

    \Xhline{1pt}
	\end{tabularx}
	\caption{
	  Ablation experiments intended to demonstrate the effectiveness of each detailed techniques for modelling two heterogeneous graphs (spectral and temporal). Performance reported in ``average(best)'' after three repeated experiments.  
	}
 	\vspace{-10pt}
	\label{tab:ablations}
\end{table}

\section{Experiments and results}
\label{sec:ExpAndResult}
\subsection{Dataset and metrics}
\label{ssec:database}
All experiments were performed using the ASVspoof 2019 logical access (LA) dataset~\cite{todisco2019asvspoof,wang2020asvspoof}.
It comprises three subsets: train, development, and evaluation. 
The train and the development sets contain attacks created from six spoofing attack algorithms (A01-A06), whereas the evaluation set contains attacks created from thirteen algorithms (A7-A19).
Readers are referred to~\cite{wang2020asvspoof} for full details.

We use two metrics: the default minimum tandem detection cost function (min t-DCF)~\cite{Kinnunen2018tdcf} and the equal error rate (EER). 
The min t-DCF shows the impact of spoofing and the spoofing detection system upon the performance of an automatic speaker verification system whereas the EER  
reflects purely standalone spoofing detection performance. 

Wang et al.~\cite{wang21fa_interspeech} showed that the performance of spoofing detection systems can vary significantly with different random seeds. 
We observed the same phenomenon; when trained with different random seeds, the EER of the baseline  RawGAT-ST~\cite{tak2021end} system was found to vary between 1.19\% and 2.06\%. 
Thus, the results are slightly different from the reported ones in~\cite{tak2021end}. 
All results reported in this paper are average results in addition to the best result from three runs with different random seeds. 

\subsection{Implementation details}
\label{ssec:implementation}
AASIST was implemented using PyTorch, a deep learning toolkit in Python. 
Inputs in the form of raw waveforms of $64,600$ samples ($\approx 4$ seconds) are fed to the RawNet2-based encoder. 
The first layer of the encoder, sinc-convolution~\cite{ravanelli2018speaker}, has $70$ filters. 
The RawNet2-based encoder consists of six residual blocks. 
The first two have $32$ filters while the remaining four have $64$ filters. 
The first two graph attention layers have $64$ filters. 
Graph pooling layers remove $50\%$ and $30\%$ of spectral and temporal nodes, respectively. 
All subsequent graph attention layers have $32$ filters and are followed by graph pooling which further reduces the number of nodes by 50\%.
We used Adam optimizer~\cite{Kingma2015adam} with a learning rate of $10^{-4}$  and cosine annealing learning rate decay. 
The AASIST-L system was tuned with a population-based training algorithm over 7 generations and with 30 experiments for each generation~\cite{jaderberg2017population}.

\subsection{Results}
\label{ssec:result}
Table~\ref{tab:breakdown} describes the EERs for each individual attacks, pooled min t-DCF, and pooled EER. 
For pooled performances, we also report the best performance inside brackets. 
It shows a performance comparison for the proposed AASIST model and the state-of-the-art baseline RawGAT-ST~\cite{tak2021end} model. 
ASSIST performs similarly or better than the baseline for 9 of the 13 conditions.  For the remaining 4 conditions for which the baseline performs better, the differences are modest.  For conditions where AASIST performs better, improvements can be substantial, e.g.\ for the A15 condition where AASIST outperforms the baseline by over 35\% relative (1.03\% vs 0.65\%).
Pooled min t-DCF and EER results are shown in the two right-most columns of Table~\ref{tab:breakdown}. AASIST outperforms the RawGAT-ST baseline in terms of both the pooled min t-DCF and the pooled EER.  For AASIST, the min t-DCF drops by over 20\% relative (0.0443 vs 0.0347).
With the best performing seed, AASIST demonstrates an EER of 0.83\% and a min t-DCF of 0.0275, beyond that of any reported single systems in the literature. 

\newpara{Comparison with state-of-the-art systems.}
Table~\ref{tab:sotaNsmall} presents a comparison of the proposed AASIST model to the performance of a number of recently proposed competing systems~\cite{tak2021graph,tak2021end,wang21fa_interspeech,zhang2021effect,hua2021towards,ge2021raw,li2021channelwise,chen2020generalization,luo2021capsule,zhang2021one,li2021replay,ma2021improved,tak2020spoofing,ge2021} for the same ASVspoof 2019 LA dataset. 
The set of systems covers a broad range of different front-end representations and model architectures.
Five of the top six systems operate upon raw waveform inputs while the top three systems are based upon graph attention networks. 
The proposed AASIST system is the best performing of all.

\newpara{AASIST-L: lightweight variant.}
Table~\ref{tab:sotaNsmall} also shows a comparison in terms of complexity for the smaller AASIST-L model and other models for which the number of parameters is openly available. 
In using only $85$K parameters, AASIST-L is substantially less complex than all other systems.
The min t-DCF and EER achieved by the AASIST-L model are better than those of all other systems except for the full AASIST model.
If appropriately modified using techniques such as half-precision inference and parameter pruning, we believe that it would be small enough to be used in embedded systems. 

\newpara{Ablations.}
Table~\ref{tab:ablations} shows results for ablation experiments for which one of the components in the AASIST model is removed. 
Results show that all three techniques are beneficial; without any one of them, results are worse than the full AASIST model.
Applying heterogeneous attention had the most impact in terms of performance. 
MGO, a similar concept with the max feature map~\cite{wu2018light} was also effective, showing consistent impact of max operation in spoofing detection.

\section{Conclusion}
\label{sec:conclusion}
We propose AASIST, a new end-to-end spoofing detection system based upon graph neural networks. 
New contributions are threefold: (i) a heterogeneous stacking graph attention layer (HS-GAL) used to model spectral and temporal sub-graphs consisted of a heterogeneous attention mechanism and a stack node to accumulate heterogeneous information; (ii) a max graph operation (MGO) that involves a competitive selection of artefacts; (iii) a modified readout scheme. 
AASIST improves upon the performance of state-of-the-art baseline by over 20\% relative in terms of min t-DCF. 
Even the lightweight version, AASIST-L, with 85K parameters, outperforms all competing systems. 

\clearpage
\bibliographystyle{IEEEbib}
\bibliography{refs}
\end{document}